\documentclass[11pt]{book}
\usepackage[top=0.5in,bottom=0.7in,right=0.7in,left=0.7in]{geometry}
\usepackage{amsmath} 
\usepackage{amssymb} 
\usepackage{tensor}
\usepackage{float}

\usepackage{fancyhdr}
\usepackage{multicol}
\usepackage{wrapfig}
\usepackage{memhfixc}
\usepackage{titlesec}
\usepackage{amsbsy}

\usepackage[T2A]{fontenc}
\usepackage[utf8]{inputenc}
\usepackage[russian,english]{babel}

\usepackage{setspace}
\usepackage{makeidx} 
\usepackage[usenames,dvipsnames]{xcolor}
\definecolor{darkblue}{RGB}{0,50,100} 
\usepackage{hyperref}
\hypersetup{
	colorlinks,
	citecolor=black, 
	linkcolor=black, 
	urlcolor=Black}

\usepackage[T2A]{fontenc}

\usepackage{appendix}

\usepackage[euler]{textgreek}

\newfont{\chapterNumber}{pplr scaled 5000}
\titleformat{\chapter}[block]%
{\relax}%
{\hspace{20pt} \llap{\color{gray}\chapterNumber\thechapter%
	}  
}{10pt}%
{\raggedright\LARGE\textsc}

\usepackage{array}
\newcolumntype{M}[1]{>{\centering\arraybackslash}m{#1}}
\newcolumntype{N}{@{}m{0pt}@{}}

\usepackage{textcomp}

\usepackage{chngcntr}
\counterwithout{equation}{chapter}
\counterwithout{figure}{chapter}
\counterwithout{section}{chapter}
\counterwithin{equation}{section}

\usepackage{array}
\usepackage{tabulary}
\newcolumntype{K}[1]{>{\centering\arraybackslash}p{#1}}

\usepackage[sectionbib]{chapterbib} 

\usepackage{fancyhdr}
\usepackage{hang} 

\usepackage{isomath} 

\usepackage{mathtools}

\usepackage{xcolor}

\makeatletter
\def\mathcolor#1#{\@mathcolor{#1}}
\def\@mathcolor#1#2#3{%
	\protect\leavevmode
	\begingroup
	\color#1{#2}#3%
	\endgroup
}
\makeatother

\usepackage{relsize}
\usepackage{float}
\usepackage{pdfpages}
\usepackage{CJKutf8}

\usepackage{url}

\RequirePackage{graphicx}
\usepackage{enumerate}
\usepackage{bm}
\usepackage{braket}
\usepackage{slashed}
\usepackage{comment}
\usepackage{color}

\usepackage{amssymb,latexsym}
\usepackage{amsmath}
\usepackage{amsthm}
\usepackage{authblk}  
\usepackage{color}
\usepackage{tikz}
\usepackage{caption}
\usetikzlibrary{snakes,decorations.pathmorphing}
\usepackage{subcaption}
\definecolor{bello}{RGB}{82,74,208}

\usepackage{graphicx}
\usepackage{soul}
\usepackage{xcolor}
\usepackage[square,numbers,comma]{natbib}
\usepackage{setspace}
\usepackage{fullpage}

\begin{document}

\chapter[Torsion and Chern-Simons gravity  in  4D space-times from a Geometrodynamical four-form\\ \textit{Patrick Das Gupta}]{Torsion and Chern-Simons gravity  in  4D space-times from a Geometrodynamical four-form\\ \hspace{-1.2mm} {\normalsize Patrick Das Gupta}%
	\chaptermark{ }}

\setcounter{section}{0}

\vspace{-3mm}
	\begin{tabular}{cp{14cm} cp{9cm}}
		\textbf{Abstract} &
		\begin{spacing}{0.8}	
			{\footnotesize 	
In Hermann Minkowski's pioneering mathematical formulation of special relativity, the space-time geometry in any inertial frame is described by the line-element $ds^2= \eta_{\mu \nu} dx^\mu dx^\nu$. It is interesting to note that not only the Minkowski metric $\eta_{\mu \nu} $ is invariant under proper Lorentz transformations, the totally antisymmetric Levi-Civita tensor $e_{\mu \nu \alpha \beta} $ too is.

In Einstein's general relativity (GR), $\eta_{\mu \nu} $ of the flat space-time gets generalized   to a dynamical, space-time dependent metric tensor $ g_{\mu \nu}  $ that characterizes a curved space-time geometry. In the present study, it is put forward that the flat space-time Levi-Civita tensor gets elevated to a dynamical four-form field $\tilde {w} $ in curved space-time manifolds,  i.e. $e_{\mu \nu \alpha \beta} \rightarrow w_{\mu \nu \alpha \beta} (x) = \phi (x) \  e_{\mu \nu \alpha \beta} $, so that  $\tilde {w} = {1\over {4!}} \ w_{\mu \nu \rho \sigma} \  \tilde{d} x^\mu \wedge   \tilde{d} x^\nu \wedge   \tilde{d} x^\rho \wedge  \tilde{d} x^\sigma$. It is  shown  that this geometrodynamical four-form field   extends GR by leading  naturally  to a torsion in the theory  as well as  to a Chern-Simons gravity.   Furthermore, it  is  demonstrated that the  scalar-density $\phi (x)$ associated with the geometrodynamical four-form $\tilde {w} $ may be used to  construct a  generalized exterior derivative that converts a p-form density to  a (p+1)-form density of  identical weight.  


In order to relate the hypothesized  four-form field $\tilde {w}$ to observational evidence, we first argue that  the associated scalar-density $\phi (x)$  corresponds to an axion-like  pseudo-scalar field  in the Minkowski space-time, and that it can also masquerade as  dark matter. Thereafter, we provide a simple semi-classical analysis in which a self-gravitating Bose-Einstein condensate  of such ultra-light pseudo-scalars  leads to the formation of a  supermassive black hole.  
A brief analysis of propagation of weak gravitational waves in the presence of $\tilde{w} $ is also considered in this article.  			
			}
		\end{spacing}
	\end{tabular}
	
	\vspace{-5mm}
	
	\noindent {\footnotesize \textit{Keywords}: Torsion, Chern-Simons gravity, Dynamical four-form, Ultra-light Dark Matter, Supermassive Black holes, Bose-Einstein Condensation}
	
	\vspace{7mm}

\section{Introduction}

By envisaging a four dimensional  manifold's space-time geometry be described by a line-element, $ds^2= \eta_{\mu \nu} dx^\mu dx^\nu$, Hermann Minkowski had overturned the idea of treating space and time disjointly. Furthermore, the cause and effect relation, in the realm of classical  physics, attained a significant elucidation in the  Minkowskian framework: infinitesimally separated events (assuming a signature (+ - - -)) for which $ds^2 < 0 $  ($ds^2 \geq 0$) are causally disconnected (causally connected).  Of course, the underlying physics involved in it is that an effect essentially follows a cause due to transfer of energy from the latter to the former, and that the rate of flow of energy is limited by the speed $c$.  What is  usually not stressed is that, because of this causality criteria,  the imaginary numbers entered physics in a fundamental manner even before the advent of quantum mechanics, i.e. the proper distance $ds$ between any two causally disconnected events is always an imaginary number.

 Although   space and time coordinates get mixed up while transiting  from one inertial frame to another, the Minkowski metric  $\eta_{\mu \nu} $ itself remains invariant.  This is utilized in  the Minkowskian framework to define an arbitrary Lorentz transformation $x^\mu \rightarrow x^{' \mu}= \Lambda^\mu_{\ \nu} x^\nu $ by the condition,
 \begin{equation}
 \eta_{\mu \nu}  = \Lambda^\alpha_{\ \mu} \Lambda^\beta_{\ \nu}  \ \eta_{\alpha \beta}
 \end{equation}

It must be pointed out that there is another tensor in the flat space-time that is invariant under proper Lorentz transformations: the totally antisymmetric Levi-Civita tensor $\epsilon_{\mu \nu \rho \sigma}$. 
Since, under $x^\mu \rightarrow  \Lambda^\mu_{\ \nu} x^\nu $,
\begin{equation}
\epsilon_{\mu \nu \alpha \beta} \ \ \ \ \rightarrow \ \  \Lambda^\sigma_{\ \mu} \Lambda^\tau_{\ \nu}   \Lambda^\gamma_{\ \alpha} \Lambda^\kappa_{\ \beta}  \ \epsilon_{\sigma \tau \gamma \kappa}
\end{equation}
\begin{equation}
=det(\Lambda)\ \epsilon_{\mu \nu \alpha \beta} \ \ ,
\end{equation}
and $det(\Lambda)=1$ for proper  Lorentz transformations, it is clear that the flat space-time Levi-Civita tensor does not change under arbitrary boosts and rotations of inertial frames.

Given this special status that the Minkowski metric and the Levi-Civita tensor enjoy in the Minkowskian space-time, as well as the fact that $\eta_{\mu \nu}$ 
  metamorphoses into a dynamical space-time metric $g_{\mu \nu}(x) $ in GR, raises a pertinent point of   elevating $\epsilon_{\mu \nu \alpha \beta}$ to a dynamical field. In the present article, we explore such a possibility by positing a  geometrodynamical field,  $w_{\mu \nu \rho \sigma} (x) $, totally antisymmetric in its indices,  that is independent of the metric tensor and that influences space-time physics. Then, the interesting feature that  $\eta_{\mu \nu}$ and $\epsilon_{\mu \nu \alpha \beta}$ both share in the flat space-time becomes relevant also  in arbitrary curved space-times \cite{pdg-ref-20}. 
  
 We propose  that this geometrodynamical field represents a new physical degree of freedom  that couples universally to all matter. In the following section,  we discuss  some of the differential geometric  aspects of $w_{\mu \nu \rho \sigma} (x) $ that may be significant in the context of space-time physics.
  
\section{Geometrodynamical four-form field and Torsion}

Now, in any 4D space-time manifold,  because of its complete antisymmetry, the four-form field $w_{\mu \nu \rho \sigma} (x)$ has only one algebraically independent component and hence, renders itself to be expressed as,
\begin{equation}
w_{\mu \nu \rho \sigma} (x) = \phi (x) \ \epsilon_{\mu \nu \rho \sigma} \ \ ,
\end{equation} 
 where $\phi (x) $ is a scalar-density of weight +1, with  $\phi (x) \rightarrow \phi^\prime (x^\prime) = \phi (x)  /J (x,x^\prime)$ under a general coordinate transformation $x^\mu \rightarrow x^{\prime \mu}$, $J(x,x^\prime)$ being the   corresponding  Jacobian.
 
In a coordinate basis, this four-form field  may be expressed as,
\begin{equation}
\tilde {w} = {1\over {4!}}\  w_{\mu \nu \rho \sigma} (x) \ 
 \tilde{d} x^\mu \wedge   \tilde{d} x^\nu \wedge   \tilde{d} x^\rho \wedge  \tilde{d} x^\sigma 
\end{equation}
\begin{equation}
= {1\over {4!}}\ \phi (x)\ \epsilon_{\mu \nu \rho \sigma} \ \tilde{d} x^\mu \wedge   \tilde{d} x^\nu \wedge   \tilde{d} x^\rho \wedge  \tilde{d} x^\sigma 
\end{equation}
Eq.(2.3) reflects the equivalence between four-forms and 0-forms in the case of four dimensional manifolds.  

Since we demand  that this new dynamical field  be independent of the metric tensor, instead of raising the indices of $ w_{\mu \nu \rho \sigma} $ 
 using the metric tensor, the totally antisymmetric components $w^{\mu \nu \rho \sigma} (x) $ corresponding to the four-form field $\tilde {w}$ are obtained from  the associated  p-vector {\bf w}, so that one obtains the relation,  $w^{\mu \nu \rho \sigma} \ w_{\mu \nu \rho \sigma}= - 4!$ \cite{Schutz80}. This implies,
\begin{equation}
w^{\mu \nu \rho \sigma} (x) = \epsilon^{\mu \nu \rho \sigma}/\phi(x) \  \ .
\end{equation}
  
We assume throughout that $\phi (x) = w_{ 0 1 2 3} (x) $ has no physical  dimension. As the canonical volume-form in GR is given by,
\begin{equation}
\tilde {V} = {1\over {4!}} \sqrt{-g}\ \  \epsilon_{\mu \nu \rho \sigma} \  \tilde{d} x^\mu \wedge   \tilde{d} x^\nu \wedge   \tilde{d} x^\rho \wedge  \tilde{d} x^\sigma \ ,
\end{equation} 
the geometrodynamical four-form field of eq.(2.3) may be expressed in terms of a scalar field $\chi (x)$,
\begin{equation}
\chi (x) \equiv \frac {\phi (x}  {\sqrt{-g}} \ \ \Rightarrow \ \ \tilde {w} = \chi (x) \tilde {V}
\end{equation}
A fundamental four-form field  that varies with space and time may be utilized to extend GR by 
 having a  dynamical torsion field in the theory. Introducing  an asymmetric affine connection (e.g. see  \cite{Hehl, Hehl1}),
\begin{equation}
\bar {\Gamma} ^\mu _{\alpha \beta} \equiv \Gamma ^\mu _{\alpha \beta} - a_T \ g^{\mu \nu} g^{\gamma \lambda} \frac {\partial \chi} {\partial x^\lambda} \ w_{\nu \gamma \alpha \beta} =  \Gamma ^\mu _{\alpha \beta} - a_T \ g^{\mu \nu} \chi^{; \gamma} \ w_{\nu \gamma \alpha \beta}  \ ,
\end{equation}
where $ \Gamma^\mu _{\alpha \beta} $ is the standard Christoffel-Levi Civita connection  while $a_T$ is a dimensionless constant that is a measure of the coupling between the metric tensor and the four-form,  we define  a  torsion  field,
\begin{equation}
 S^{\ \ \ \mu}_{\alpha \beta} (x) =  \frac {1}{2} \bigg (\bar {\Gamma} ^\mu _{\alpha \beta} - \bar {\Gamma} ^\mu _{\beta \alpha } \bigg ) = - a_T \ g^{\mu \nu} \chi^{; \gamma}\ w_{\nu \gamma \alpha \beta}=  - a_T \sqrt{-g} \ g^{\mu \nu} g^{\gamma \sigma}\ \chi \ \frac{\partial \chi}{\partial x^\sigma} \ \epsilon_{\nu \gamma \alpha \beta}
\end{equation}

In the language of differential geometry, since torsion is a vector valued 2-form, the above expression for it can be seen as arising from an inner contraction (e.g. see \cite{Mukunda, Schutz80}) of $\tilde {w}$ with the vector field $\chi^{; \gamma} \frac {\partial} {\partial x^\gamma}$ 
followed by inner contractions with four vector fields $- g^{\mu \nu} \frac {\partial} {\partial x^\nu}$, $\mu=0,1,..3$. 

Covariant derivatives of tensor fields are obtained in  the usual way,
\begin{equation}
D_\nu A^\mu = \partial_\nu A^\mu + \bar {\Gamma} ^\mu _{\nu \beta} A^\beta = A^\mu _{\ ;\nu}- a_T \  g^{\mu \alpha} \chi^{; \gamma} w_{\alpha \gamma \nu \beta}  A^\beta \ \ ,
\end{equation}
\begin{equation}
D_\nu B_\mu = \partial_\nu B_\mu - \bar {\Gamma} ^\beta _{\mu \nu } B_\beta = B_{\mu ;\ \nu} + a_T  g^{\beta \sigma} \chi^{; \gamma} w_{\sigma \gamma \mu \nu }  B_\beta \ \ ,
\end{equation}
and,
\begin{equation}
D_\nu C_{\mu \tau}= \partial_\nu C_{\mu \tau}  - \bar {\Gamma} ^\beta _{\mu \nu } C_{\beta \tau} -  \bar {\Gamma} ^\beta _{\tau \nu } C_{\mu \beta } = C_{\mu \tau;\ \nu} + a_T g^{\beta \sigma} \chi^{; \gamma} [w_{\sigma \gamma \mu \nu }  C_{\beta \tau}  +  w_{\sigma \gamma \tau \nu }  C_{\mu \beta}] \ \ ,
\end{equation}
with $C_{\mu \tau;\ \nu}$ being the covariant derivative of $ C_{\mu \tau}$ in standard GR when there is no torsion (i.e. $a_T=0$ case). Using eq.(2.11), the metric compatibility of $S^{\ \ \ \mu}_{\alpha \beta}$ can easily be proved by showing that $D_\alpha g_{\mu \nu}=0 $. Similarly, for any vector field $A^\alpha (x) $, eq.(2.7) entails that,
  $$\bar {\Gamma} ^\mu _{\alpha \beta} A^\alpha A^\beta = {\Gamma} ^\mu _{\alpha \beta} A^\alpha A^\beta \ $$
implying that the torsion field given by eq.(2.8)  is autoparallel compatible too.
In our case, the contortion tensor $K^{\ \ \mu}_{\alpha \beta}$ and $S_{\alpha \beta \gamma} $  have the expressions,
 \begin{equation}
 K^{\ \ \ \mu}_{\alpha \beta} \equiv  - S^{\ \ \ \mu}_{\alpha \beta} + S^{\ \mu}_{\beta \  \alpha} - S^\mu_{\ \ \alpha \beta} = - S^{\ \ \ \mu}_{\alpha \beta}
 \end{equation}
and,
\begin{equation}
S_{\alpha \beta \gamma} = g_{\gamma \mu}  S^{\ \ \ \mu}_{\alpha \beta} = a_T \ \chi^ {;\sigma } w_{\sigma \alpha \beta \gamma}= a_T \sqrt{-g} \  g^{\gamma \sigma} \chi \ \frac{\partial \chi}{\partial x^\gamma} \epsilon_{\sigma \alpha \beta \gamma}
\end{equation}
which is simply an inner contraction of  $\tilde {w} $ with the vector field $\chi^ {;\sigma } \frac {\partial} {\partial x^\sigma} $.

The important point to note is that the torsion vanishes if either $a_T =0$ or if the scalar field $\chi (x)$ that  corresponds to the geometrodynamical four form is independent of both space as well as  time. {\bf  For the sake of simplicity,  in the remaining portions of the article we will concentrate on the   case   $a_T =0$ while assuming $\frac{\partial \chi}{\partial x^\alpha} $ to be non-vanishing}. It is  straightforward to extend the analysis to the $a_T \neq 0$ case in order to study the effects of torsion. For the rest of the sections, we have chosen a unit system in which $\hbar=c=1$ and the metric signature:  (+ - - -).

\section{Action, Field Equations  and Chern-Simons extensions} 
\vskip 1.5 em
\subsection{Dynamical equations}
\vskip 1.5 em
As $\phi $ is a scalar-density of weight +1,
\begin{equation}
\phi_{\ ; \beta} =  \phi_{\ , \beta} - \Gamma^\alpha_{\alpha \beta}\ \phi  \ \ ,
\end{equation}
where we follow the standard notation,
\begin{equation}
f_{\ , \beta}  \equiv \partial_\beta f  \ \ .
\end{equation}

Using the above equations along with the relation 
 $w^{\mu \nu \rho \sigma} \ w_{\mu \nu \rho \sigma}= -\ 4!$, the covariant derivatives of the four-form field and its corresponding p-vector are easily shown to be,
\begin{equation}
w_{\mu \nu \alpha \beta \ ; \lambda} = [(\ln \phi)_{, \lambda} - \Gamma ^\sigma _{\sigma \lambda}] w_{\mu \nu \alpha \beta} \ \ ,
\end{equation}
and,
\begin{equation}
w^{\mu \nu \alpha \beta}_{\ \ ; \lambda} = - [(\ln \phi)_{, \lambda} - \Gamma ^\sigma _{\sigma \lambda}] w^{\mu \nu \alpha \beta} \ \ ,
\end{equation}
respectively.

The action $\mathcal{A}$ for the standard matter, geometry as well as the dynamical four-form that is invariant under general coordinate transformations may be expressed as,
\begin{equation*}
   \mathcal{A} = -  {{m^2_{Pl}}\over{16 \pi}} \int {R \sqrt {-g} \ d^4x}+ \int {L \sqrt {-g}\ d^4 x}\  +  
\end{equation*}
\begin{equation}
  \ \ \ \ \ \ + \ {{a_1}\over{4!}} \int {\phi \ w^{\mu \nu \alpha \beta}_{\ \ ;\lambda}  w_{\mu \nu \alpha \beta}\ ^{;\lambda} \ d^4 x} + a_2 \int {\phi (x) d^4 x} \ ,
\end{equation}

where $m_{Pl}\equiv \sqrt{\hbar c/G}$ and $L$ are the Planck mass and the Lagrangian density of the matter fields, respectively, with  $a_1$ and $a_2$ being real valued constants  of the theory having physical dimensions $(\mbox{mass})^2$ and $(\mbox{mass})^4$, respectively.  The part of the
action in eq.(3.5) that pertains to the four-form is by no means unique. For instance,  a term $\propto  \int {R \ \phi \ d^4x}$ could be added to the above action, but   we limit ourselves to gravitational minimal coupling, at present. Later, we
shall discuss  supplementing the above action with Chern-Simons terms induced by  the geometrodynamical four-form.

By extremising $\mathcal{A}$  with respect to $g_{\mu \nu}$ and $\phi $, respectively, we obtain the following equations of  motion,
\begin{equation}
 R_{\mu \nu} - {1\over{2}} g_{\mu \nu} R = {{8 \pi}\over{m^2_{Pl}}} [T_{\mu \nu} + \Theta _{\mu \nu}]  \ , 
\end{equation}
\begin{equation}
    \chi ^{\ ; \alpha}_{\ ; \alpha} \equiv {1\over{\sqrt {-g}}} (\sqrt {-g} g^{\alpha \beta} \chi_{ , \alpha})_{ , \beta} = {1\over {2}} \bigg [ g^{\mu \nu} {{\chi_{,\mu } \chi_{,\nu}}\over{\chi}} + {a_2\over {a_1}} \chi \bigg ] \ ,
\end{equation}

where the scalar field $\chi (x) $  is defined, as earlier,  to be $\chi \equiv {\phi \over {\sqrt{-g}}} $, while $T_{\mu \nu}$   and  $\Theta _{\mu \nu}$ are the energy-momentum tensors for the standard matter and the geometrodynamical four-form, respectively.  The expression for the latter is given by,
\begin{equation}
 \Theta _{\mu \nu} = 2 a_1 \bigg [{{\chi_{,\mu } \chi_{,\nu}}\over{\chi}} - g_{\mu \nu} \chi ^{\ ; \alpha}_{\ ; \alpha} \bigg ] \ .   
\end{equation}
When $\chi $ satisfies the equation of motion given by eq.(3.7), the above energy-momentum tensor gets simplified to,
\begin{equation}
\Theta _{\mu \nu} = 2 a_1 \bigg [{{\chi_{,\mu } \chi_{,\nu}}\over{\psi}} - {{g_{\mu \nu}}\over {2}} \bigg ( g^{\alpha \beta} {{\chi_{, \alpha } \chi_{,\beta}}\over{\chi}} + {a_2\over {a_1}} \chi \bigg )  \bigg] \ .    
\end{equation}
\vskip 1.5 em
\subsection{Chern-Simons extensions}

In this sub-section, we  construct Chern-Simons (CS) terms in (3+1)-dimensional  space-time.  Our  CS extensions of GR follows closely the seminal paper of  Jackiw and Pi \cite{Jackiw}, except for a crucial difference: they had introduced an external fixed four-vector $v_{\mu}$ instead of a dynamical $\phi _{;\mu}$ in their formulation. Because of this, their model had a manifest  violation of Lorentz invariance.  In our case, both general as well as Lorentz covariances are maintained throughout \cite {pdg-ref-20}.  

 First, we consider  a Chern-Simons (CS)  term that couples the electromagnetic field to the four-form $\tilde{w} $,
 \begin{equation*}
   \mathcal{A}_{CS}= \mbox{J}\int {w^{\mu \nu \alpha \beta} F_{\mu \nu} A_\alpha \phi_{;\beta} \ d^4 x}   
 \end{equation*}
\begin{equation}
  \ \ \ =\mbox {J} \int {\epsilon^{\mu \nu \alpha \beta} F_{\mu \nu} A_\alpha (\ln \chi)_{,\beta} \ d^4 x}\ ,  
\end{equation}


where J is a  dimensionless constant and  $F_{\mu \nu} = A_{\nu, \mu } - A_{\mu , \nu}$. 
The action expressed in eq.(3.10)  is invariant under diffeomorphism as well as  gauge transformations, $A_\mu \rightarrow A_\mu + \partial_\mu \xi (x) $.

The upshot of adding $\mathcal{A}_{CS}$  to the standard action $\mathcal{A}_{EM}$  for an electromagnetic field interacting with charge particles in the presence of gravitation  is that, upon extremising the full action $\mathcal{A}_{EM} + \mathcal{A}_{CS}$ with respect to $A_\mu$, one obtains the   following modified Einstein-Maxwell equation,
\begin{equation}
F^{\alpha \beta}_{\ \ ; \beta}= -4 \pi j^\alpha + 8\pi \mbox{J}\  w^{\mu \nu \alpha \beta} F_{\mu \nu} \ \chi_{,\beta} \ ,    
\end{equation}
 with $j^\alpha$ being the 4-current density associated with charge particles. 

 Use of  the covariant derivative to the above equation with respect to $x^\alpha $ leads to the standard 
 charge density continuity equation that entails conservation of electric charge $Q$,
 \begin{equation}
      (\sqrt {-g} j^\alpha)_{, \alpha}=0 \ \ \Rightarrow \ \ Q=\int{j^0 \sqrt{-g} \ d^3x}\ \  .    
 \end{equation}

Application of the above  CS formulation to study its  effects on magnetohydrodynamics as well as the $\vec{E}.\vec{B} \neq 0$ regions of a pulsar magnetosphere acting as a  source of propagating $\tilde{w} $ have been studied earlier \cite{pdg-ref-47}. 

\vskip 1.5 em
We continue adopting the procedure laid out in the paper by Jackiw and Pi  \cite{Jackiw} to construct a Chern-Simons action for gravity in the 3+1-dimensional space-time. Instead of their fixed Lorentz vector $v_\mu$, we use the covariant derivative of $\phi (x) $ so that,
\begin{equation*}
    \mathcal{A}_{GCS}= \mbox{H}\int {w^{\mu \nu \alpha \beta} [\Gamma^\sigma_{\nu \tau} \partial _\alpha \Gamma^\tau_{\beta \sigma} + {2\over{3}} \Gamma^\sigma_{\nu \tau}\Gamma^\tau_{\alpha \eta}\Gamma^\eta_{\beta \sigma}]\phi_{;\mu} \ d^4 x}
\end{equation*}
\begin{equation}
  \ \ \ =\mbox {H} \int {\epsilon^{\mu \nu \alpha \beta} [ \Gamma^\sigma_{\nu \tau} \partial _\alpha \Gamma^\tau_{\beta \sigma} + {2\over{3}} \Gamma^\sigma_{\nu \tau}\Gamma^\tau_{\alpha \eta}\Gamma^\eta_{\beta \sigma}](\ln {\chi})_{,\mu} \ d^4 x} \ \ \ ,  
\end{equation}
where H is a dimensionless, real constant. After integrating by parts once, eq.(3.13) can be expressed in terms of the Riemann tensor as,
\begin{equation}
  \mathcal{A}_{GCS}= -{\mbox{H}\over{2}} \int {\ln {\chi}\ {}^*R R \ d^4 x} \ \ ,   
\end{equation}
with,
\begin{equation}
   {}^*R R \equiv {1\over{2}} \epsilon^{\mu \nu \alpha \beta}  R^\tau_{\ \sigma \alpha \beta}  R^\sigma_{\ \tau \mu \nu} = 8 \bigg [ R^{\tau \sigma}_ {\ \  01}  R_{\sigma  \tau  23} + R^{\tau \sigma}_ {\ \  12}  R_{\sigma  \tau  03}
+  R^{\tau \sigma}_ {\ \  13}  R_{\sigma  \tau  20}\bigg ] 
\end{equation}
and,
\begin{equation}
    {}^*R^{\tau \rho \mu \nu} \equiv {1\over{2}} \epsilon^{\mu \nu \alpha \beta} R^{\tau \rho}_ {\ \ \alpha \beta} \ .
\end{equation}
We have used the following conventions pertaining to the Riemann curvature and Ricci tensors,
\begin{equation*}
 R^\tau_{\ \sigma \alpha \beta} = \partial_\alpha \Gamma^\tau_{\sigma \beta} - \partial_\beta \Gamma^\tau_{\sigma \alpha} + \Gamma^\tau_{\alpha \eta} \Gamma^\eta_{\beta \sigma} - \Gamma^\tau_{\beta \eta} 
\Gamma^\eta_{\alpha \sigma}   
\end{equation*}
and $R_{\alpha \beta} = R^\tau _{\ \alpha \tau \beta}$.
(We note that  $\ln {\chi} $ in  eq.(3.13) acts like the parameter $\theta $ in   Jackiw and Pi's paper \cite{Jackiw} in which the external vector $v_{\mu}$ is set as $\theta_{,\mu}$.)

 Adding $\mathcal{A}_{CS} + \mathcal{A}_{GCS}$ to $\mathcal{A}$ of eq.(3.5) and then extremising the full action with respect to $\phi $ and $g_{\mu \nu}$ leads to,
 \begin{equation}
   \chi ^{\ ; \alpha}_{\ ; \alpha}  = {1\over {2}} \bigg [ g^{\mu \nu} {{\chi_{,\mu } \chi_{,\nu}}\over{\chi}} + {a_2\over {a_1}} \chi  + {\mbox{J}\over{ 2 a_1}}\chi \ w^{\mu \nu \alpha \beta} F_{\mu \nu} F_{\alpha \beta}
-  {\mbox{H}\over{ 4 a_1}}\chi \  w^{\mu \nu \alpha \beta}   R^\tau_{\ \sigma \alpha \beta}  R^\sigma_{\ \tau \mu \nu} \bigg ] \ ,  
 \end{equation}
\begin{equation}
    R_{\mu \nu} - {1\over{2}} g_{\mu \nu} R = {{8 \pi}\over{m^2_{Pl}}} [T_{\mu \nu} + \Theta _{\mu \nu} + C_{\mu \nu}] \ ,
\end{equation}
where the modified Cotton tensor $C^{\mu \nu}$ is defined as,
\begin{equation}
 C^{\mu \nu} \equiv -2{\mbox{H}\over{\sqrt{-g}}} \bigg [{1\over{4}} { }^* R R g^{\mu \nu} - (\ln \chi)_{; \alpha ; \beta} \bigg ( { }^* R^{\beta \mu \alpha \nu} + { }^* R^{\beta \nu \alpha \mu} \bigg ) + (\ln \chi)_{, \alpha}
\bigg (\epsilon^{\alpha \mu \sigma \tau} R^\nu _{\ \sigma ; \tau} + \epsilon^{\alpha \nu \sigma \tau} R^\mu _{\ \sigma ; \tau} \bigg ) \bigg ] \ .    
\end{equation}
The first term in the RHS of Eq.(3.19) is new and is not present in the expression for the Cotton tensor  as delineated in \cite{Jackiw}. It appears in our work because under an infinitesimal variation $g_{\mu \nu} \rightarrow$ $g_{\mu \nu} + \delta g_{\mu \nu}$, the induced change in $(\ln \chi )_{,\mu}$ occurring in eq.(3.13) (while considering  the ensuing variation of $ \mathcal{A}_{GCS} $) is  given by,
\begin{equation}
 \delta (\ln \chi)_{,\mu} = \delta ( \phi_{; \mu} / \phi)= - \delta \Gamma^\alpha_{\alpha \mu}= -{1\over{2}} \delta ( g^{\alpha \beta} g_{\alpha \beta , \mu}) \ \ .   
\end{equation}
 
However, when the equation of motion for the dynamical four-form given by eq.(3.17) is substituted in eq.(3.8), its energy-momentum tensor  in the presence of CS-term takes the form,
\begin{equation}
    \Theta _{\mu \nu}= 2 a_1 \bigg [{{\chi_{,\mu } \chi_{,\nu}}\over{\chi}} - {{g_{\mu \nu}}\over {2}} \bigg ( g^{\alpha \beta} {{\chi_{, \alpha } \chi_{,\beta}}\over{\chi}} +
 {a_2\over {a_1}} \chi \bigg )  \bigg] + {H\over{2\sqrt{-g}}} { }^* R R g_{\mu \nu}
\end{equation}
so that when eq.(3.21) is substituted in eq.(3.18) the first term in the RHS of eq.(3.19) cancels with the last term in the RHS of eq.(3.21). In other words, the term ${ }^* R R g_{\mu \nu}$ does not appear in or contribute to the Einstein equations given by eq.(3.18).

Assuming that the geometrodynamical four-form  is  a dark energy candidate, the above dynamical equations were studied in an earlier work to address the observed  late time acceleration of the expansion rate of the  universe (\cite{pdg-ref-20}, and the references therein). Our objective, in this article,  is somewhat different since  we  propose  here that the particle associated with $\tilde{w} $ is a pseudo-scalar which acts as a cold dark matter (CDM) candidate aiding in the formation of  supermassive black holes when the universe was young. This is the subject of Section IV.

\vskip 1.5 em

\subsection{Generalized Exterior Derivative}
\vskip 1.5 em
Apart from the possibility of  torsion and Chern-Simons extensions ensuing from the four-form field  $\tilde {w}$,  there may  be another differential geometric significance of this field.  The scalar-density $\phi $ can also  be used  to have an antiderivation acting on antisymmetric tensor-densities  of arbitrary weights. 

Suppose $\tilde {\alpha} $ is a p-form density with weight $w$ such that its components  $\alpha _{\nu_1 \nu_2 .. \nu_p} $ transform to $J^{-w}(x,x^\prime) \ \alpha _{\nu_1 \nu_2 .. \nu_p }$, under a general coordinate transformation $x \rightarrow x^\prime$, $J(x,x^\prime)$ being the corresponding Jacobian.
We define a generalized exterior derivative $\tilde {d}_w$ acting on $\tilde {\alpha} $ in the following manner:
\begin{equation*}
\tilde {d}_w \tilde {\alpha} \equiv {1\over{p!}}\ \partial_ \mu \alpha _{\nu_1 \nu_2 .. \nu_p }\ \tilde {d} x^\mu \wedge \tilde {d} x^{\nu _1} \wedge ... \wedge \tilde {d} x^{\nu _p} -
\end{equation*}
\begin{equation}   
- w \ \partial_\mu (\ln \phi)\ \tilde {d} x^\mu \wedge  \tilde {\alpha} 
= \tilde {d} \tilde {\alpha}  - w \ \tilde {d}\ (\ln \phi)\ \wedge  \tilde {\alpha} \ .
\end{equation}
It is easy to verify that $\tilde {d}_w \tilde {\alpha} $ is a (p+1)-form density of weight $w$. 

Suppose $\alpha _1 $ and $\alpha_2 $   are scalar-densities of weights $w_1$ and $w_2 $, respectively. Then, from eq.(3.22), we have,
\begin{equation}
 \tilde {d}_w \alpha_i = \partial_\mu  \alpha_i \ \tilde {d} x^\mu  - w _i \alpha_i \ \partial_\mu (\ln \phi)\ \tilde {d} x^\mu  
, \ \ \ i=1,2   
\end{equation}
 which are one-form densities and furthermore, one can show that,
\begin{equation}
\tilde {d}_w (\alpha_1 \tilde {d}_w \alpha_2) =  \tilde {d}_w \alpha_1 \wedge \tilde {d}_w \alpha_2 \ .    
\end{equation} 
The generalized exterior derivative  defined by eq.(3.22) also satisfies (a) $\tilde {d}_w \tilde {d}_w =0 $ and  (b) $\tilde {d}_w (\tilde {\alpha} \wedge \tilde {\beta}) 
= \tilde {d}_w \tilde {\alpha} \wedge \tilde {\beta} + (-1)^p \tilde {\alpha } \wedge \tilde {d}_w \tilde {\beta} $, where $ \tilde {\alpha} $ and $\tilde {\beta} $ are p- and q-form densities of weights $w_1 $ and $w_2 $, respectively.
These properties are sufficient to qualify $\tilde {d}_w $ to be a well defined anti-derivation on differential form-densities \cite {Mukunda}. 
 
From eq.(3.23) it follows that,
\begin{equation}
 \tilde {d}_w \sqrt {-g}  = - \sqrt {-g} \ \tilde {d} \ln ({\phi \over {\sqrt {-g}}}) \ ,    
\end{equation}
since $\sqrt {-g} $ is a scalar-density of weight +1. 
 
 There are other physically meaningful tensor densities e.g.   Dirac delta function (which is a scalar-density)  and anti-symmetric tensor-densities, e.g. $\widetilde{F}^{\mu \nu}=$ dual of $F^{\mu \nu}$, on which  $\tilde {d}_w $  can act.
 In passing, we note that  $\tilde {d}_w \phi = 0$, which is  analogous to $g_{\mu \nu ; \lambda} = 0$. 

\vskip 1.5 em
\section{Dark bosons, Bose-Einstein Condensates  and Formation of Supermassive Black Holes}
\vskip 1.5 em
The wave-particle duality of quantum mechanics transcends to field-particle duality in quantum field theory  (QFT).  Particles associated with quantum fields are intimately linked with their Poincar$\mbox{e}^\prime$ algebra and  special relativistic covariance in the standard framework of QFT. In the context  of  the four-form field $\tilde{w}$, or equivalently the scalar-density $\phi (x) $,  we may ask what kind of particle is associated with it? 

Now, $\phi (x) $ in the  flat Minkowski space-time transforms to,
\begin{equation}
    \phi \rightarrow \phi^\prime = J(t,\vec{r} ; t,-\vec{r}) \ \phi = - \phi \ \ ,
\end{equation}
under a space reversal $\vec{r} \rightarrow - \vec{r} $ as the Jacobian $J(t,\vec{r} ; t,-\vec{r} )= -1 $. Since $\phi $ changes its sign under a parity transformation, the quantum particle associated with the   field $\tilde {w} $ is a pseudo scalar, and thus is a boson.

Because such a  pseudo-scalar is  likely to interact with matter with a strength at the most comparable with that of gravity, we propose here  that $\tilde{w} $  is the  candidate for ultra-light cold dark matter (CDM),  although such  CDM particles have been conventionally linked with axions in the literature (e.g. see \cite{Fukuyama}, and the references therein). Can the dark bosons associated with the geometrodynamical four-form sort out the intriguing problem of the very early formation of  supermassive black holes (SMBHs)?

So far,  around a million quasars have been observed, each  powered by the infall of matter into the deep gravitational potential of SMBHs weighing $\gtrsim 10^7 M_\odot$, and thereby getting overheated to turn into hot plasma \cite{pdg-ref-0, pdg-ref-1, pdg-ref-2, Volon}. Quasar J0100+2802 found at a redshift, $z=6.33$, is associated with a SMBH of mass $\approx  10^{10} \ M_\odot $, while another (J2157–3602)  at  a  redshift,  $z=4.7$, has a black hole weighing $\cong 2.4 \times 10^{10} \ M_\odot $.  Similarly, quasars J1120+0641 and J1342+0928 located at redshifts $z=7.09$ and  $z=7.54$, respectively, are associated with SMBHs with mass $\gtrsim 10^8 \ M_\odot $. These findings  pose a severe challenge to the existing models that invoke seed black holes (BHs) growing via accretion,  since  one needs  extremely  large  seed BHs with mass $\gtrsim 10^3 \ M_\odot$  at $z \gtrsim 40$ \cite {Volon, pdg-ref-11}.

Recently, using the JWST observations, a very distant SMBH with mass $\sim  10^6 \ M_\odot $ has been discovered  at a redshift of 10.6 \cite{SMBHatz=10.6}.
Of course,  even heavier SMBHs at lower redshifts, like  with mass $\cong 4 \times 10^{10} \ M_\odot $  at  the centre of Holm 15A galaxy belonging to the galaxy cluster Abell 85 have been detected \cite{Abell85}. At a redshift of 3.96, a SMBH weighing $\cong 1.7  \times 10^{10} \ M_\odot $, accreting matter at a rate one solar mass per day has been seen \cite{17BillionSolarMassBHz=3.962}.

In order to address the formation of  SMBHs  when the universe was  $\lesssim 10^9$ yrs old, we had earlier put forward  a scenario using the framework of Gross-Pitaevskii equation in which ultra heavy BHs  are created because of  gravitational contraction   of  Bose-Einstein condensates (BECs) of ultra-light bosonic CDM \cite{PDGEklavya, PDGFazlu}. In this section, we provide a simple semi-classical analysis that captures the essential physics of the problem.

 Considering a galactic DM halo of size $R_h$ that is constituted of these gravitationally bound dark bosons of mass $m$, a significant fraction of the bosons that have sufficiently low momenta $p$ would form a BEC  if their de Broglie wavelength,
 \begin{equation}
     \lambda_{DB} \sim \frac {h} {p} \gtrsim \bigg (\frac {3 N} {4 \pi R^3_h} \bigg )^{-1/3}= R_h \bigg (\frac {3 N} {4 \pi} \bigg )^{-1/3} =  R_h \bigg (\frac {3 M} {4 \pi m } \bigg )^{-1/3} \ \ ,
 \end{equation}
 where $M$ and  $N$  are the total mass of dark pseudo-scalars and  their  number, respectively, making up the BEC. Eq.(4.2) represents the condition for BEC formation as it entails that the de Broglie wavelength  of the ultra-light  pseudo-scalars be  larger than the mean separation between the neighbouring dark bosons.

In the context of a $\Lambda$CDM scenario with a positive cosmological constant present,  the energy $E_b$ of a typical dark boson  is given by \cite{PDGCosmo},
\begin{equation}
    E_b \sim \frac {p^2} {2m} 
 - \frac {G M m} {R_h} +   \frac{\Lambda \ r^2}{6}  m c^2 < 0  \ \ .
\end{equation}

The condition $E_b < 0$ follows from the fact that the DM halo is a gravitationally bound structure. Hence,
\begin{equation}
p^2 < \frac {2 G M m^2} {R_h} -  \frac{\Lambda \ r^2}{3}  m^2 c^2 \ \ .  
\end{equation}

Since a very weakly interacting dark boson can be anywhere within the DM halo, application of the Heisenberg's uncertainty principle implies that,
\begin{equation}
 \Delta p \sim p \gtrsim \frac {\hbar} {2 R_h} \ .   
\end{equation}
From eqs.(4.2) and (4.5), we have an inequality,
\begin{equation}
   \frac {h} {4 \pi p} \lesssim R_h \lesssim \frac {h} {p} \bigg (\frac {3 N} {4 \pi} \bigg )^{1/3}  
\end{equation}
 that is self-consistent since $N \gg 1$.

Using the result $p \sim \frac {\hbar} {2 R_h}$  from eq.(4.5) in  eq.(4.3), we obtain,
\begin{equation}
   E_b (R_h) \sim \frac {\hbar^2} {8 m R^2_h} - \frac {G M m} {R_h} +  \frac{\Lambda \ R^2_h}{6}  m c^2 < 0 \ \ , 
\end{equation}
that leads to a condition,
\begin{equation}
 R_h \gtrsim  \frac {\hbar^2} {8 G M m^2} \bigg (1- \frac{R^3_h \Lambda c^2}{6 G M}\bigg )^{-1} \ \ .   
\end{equation}
In the context of the $\Lambda$CDM model, the numerical value of the cosmological constant that ensues from the cosmological constant density parameter, $\Omega_{\Lambda,0} \cong 0.7$ and the Hubble parameter, $\mbox{H}_0 \cong 68$ km/s/Mpc, is $\Lambda \cong 10^{-56}\ cm^{-2}$. 

Even with a low dark matter halo mass $M\cong 10^6 \ M_\odot $, one finds $\lambda_M \equiv \Lambda c^2 / 6 GM \cong 10^{-68}\ cm^{-3}$. Hence, on galactic scales  $R_h \lesssim 10^{22}\ cm$, $R^3_h \Lambda c^2/ 6 G M \ll 1$ so that  eq.(4.8) can be approximated to,
\begin{equation}
    R_h \gtrsim  \frac {\hbar^2} {8 G M m^2} \bigg (1 +  \frac{R^3_h \Lambda c^2}{6 G M}\bigg ) \approx \frac {\hbar^2} {8 G M m^2} \ \ .
\end{equation}

In a BEC, most of the identical bosons occupy the ground state. Therefore,  minimizing $E_b$ (given by eq.(4.7)) with respect to $R_h$,  one obtains,
\begin{equation}
   \frac{\partial E_b}{\partial R_h}= \frac{\Lambda \ R_h}{3}  m c^2 + \frac{G M m}{R^2_h}  -  \frac {\hbar^2}{4  m R^3_h} =0 \ , 
\end{equation}  
which leads to a quartic equation that needs to be solved in order to estimate the BEC size,
\begin{equation}
 \Lambda R^4_h + \frac{3 G M }{c^2} R_h -  \frac {3 \hbar^2}{4  m^2 c^2} = 0 \  \ .   
\end{equation}
The above equation can be solved exactly and, out of the four roots, only one happens to be  real and positive. However, given the smallness of the quantity  $\lambda_M $, 
 this real and positive solution to eq.(4.11), even up to $\lambda^3_M $ orders,  is of the form, 
\begin{equation}
R_{bec} \cong \frac {\hbar^2}{4 G M m^2} \cong 22   \bigg ( \frac {10^9 M_\odot}{M} \bigg ) \bigg ( \frac {10^{-22} \ \mbox{eV}}{m} \bigg )^2 \ \mbox{pc}  \ \ ,    
\end{equation} 
 that corresponds  to a single boson energy,
\begin{equation}
 E_{min} = E_b(R_{bec}) \cong  - \frac {G M m} {2 R_{bec}} = - 2 \ m c^2  \bigg ( \frac {m^2_{Pl}}{m\ M} \bigg )^{-2}  \ .   
\end{equation} 
Eqs.(4.12) and (4.13)  are of course exact solutions for the  $\Lambda =0$ case, ensuing trivially from eq.(4.11). 

The preceding equations entail that for larger  BEC mass $M$ or larger rest mass $m$ of the pseudo-scalar not only the BEC size is smaller, the condensate  is also more tightly bound since its energy is  more negative. The physical repercussion of these features is that for a sufficiently large mass $M$ or $m$, $R_{bec} $ can be very close to the corresponding Schwarzschild radius  $R_s \equiv 2 G M/c^2$ so that even a slight perturbation would cause an irreversible gravitational collapse leading to the formation of a BH.

The BEC would form a BH  if its size $R_{bec}$ shrinks below the Schwarzschild radius $R_S$ limit,
\begin{equation}
 R_S=\frac {2 G M} {c^2} = \frac {2 M} {m^2_{Pl}}  \ \ ,   
\end{equation}
where $ m_{Pl} $ represents the Planck mass.

Using  the criteria $R_{bec} \lesssim R_S$ along with eqs.(4.12) and (4.14),  the condition for the BEC to implode into a BH is given by the inequality,
\begin{equation}
\bigg (\frac{ m^2_{Pl}} {4 M \ m} \bigg )^2 \lesssim 1 \ ,    
\end{equation}
\begin{equation}
 \Rightarrow \  m \ M \gtrsim  \ 0.25 \ m^2_{Pl}\ \Rightarrow \ M \gtrsim 3 \times 10^9 \ \bigg (\frac{ m} {10^{-20}\ \mbox{eV}} \bigg )^{-1} M_\odot \ .   
\end{equation}
A more  rigorous analysis carried out earlier, for the $\Lambda = 0$ case, by  employing  a  Gross-Pitaevskii equation framework had shown that the  dynamical evolution of ultra-light dark bosons  in the BEC phase  leads to the   formation of a BH on time scales $\sim 10^8$ years  with   a similar constraint given by eq.(4.16) \cite{PDGEklavya, PDGFazlu}.

Now, the Hawking temperature $T_{BH}$  of a BH of mass $M$  is given by \cite{Hawking},
\begin{equation}
    k_B T_{BH} = \frac{m^2_{Pl} }{8 \pi M} \ \ .
\end{equation}
So, it is interesting to note that when eq.(4.16) is substituted in eq.(4.17), one gets,
\begin{equation}
k_B T_{BH} \lesssim \frac{m}{2\pi}\ \ .    
\end{equation}
In other words,  a SMBH created from the collapse of a BEC of ultra-light dark bosons would be predominantly radiating, via Hawking evaporation, massless particles as well as these pseudo-scalars associated with $\tilde {w}$. 

\vskip 1.5 em
\section{Propagation of Gravitational Waves in the presence of the geometrodynamical four-form}
\vskip 1.5 em   
 W\lowercase{E TAKE UP IN THIS SECTION  THE EFFECT OF $\tilde{w}$ ON THE PROPAGATION OF GRAVITATIONAL WAVES} (GWs) \lowercase{ ASSUMING THAT, EXCEPT FOR A WEAK } GW and the geometrodynamical four form,
\lowercase{THERE IS NO OTHER MATTER PRESENT.} 

By making the weak field approximation $g_{\mu \nu} = \eta_{\mu \nu} + h_{\mu \nu}$ (with $\vert h_{\mu \nu} \vert \ll 1)$) and considering a traceless-transverse (TT)-gauge for a  plane
GW travelling along the x-axis, we express the GW amplitudes,
\begin{equation}
 h_{\oplus}\equiv h_{22} (t,x)= - h_{33} (t,x) \ \ \ \mbox{and} \ \ \ h_{\otimes}\equiv h_{23} (t,x)=  h_{32} (t,x) \  . \end{equation} 
R\lowercase{ETAINING ONLY UPTO QUADRATIC TERMS IN $ h_{\oplus} $ AND $ h_{\otimes}$ FOR THE GRAVITATIONAL PART}, the action given by  eqs.(3.5) and (3.14) reduce to (also, see \cite{Alex}), 
\begin{equation}
 \mathcal{A} \approx - {{m^2_{Pl}}\over{8 \pi}} \int {[h_{\oplus}  (\ddot{h}_{\oplus} - h^{\prime \prime}_{\oplus} ) +  h_{\otimes}  (\ddot{h}_{\otimes} - h^{\prime \prime}_{\otimes} )]d^4x} +
\mathcal{A}_{GCS} + \mathcal{A}_\phi    
\end{equation}
with,
\begin{equation*}
   \mathcal{A}_{GCS} \approx - H \int{\eta^{\tau \lambda} \bigg ( \frac {\partial^2 \ln \chi}{\partial x \partial x^\tau} ( h_{\oplus} h_{\otimes, 0 \lambda} - 
 h_{\otimes} h_{\oplus, 0 \lambda}) + \frac {\partial^2 \ln \chi}{\partial t \partial x^\tau} (h_{\otimes} h_{\oplus, 1 \lambda} -  h_{\oplus} h_{\otimes, 1\lambda})
 \bigg )d^4x}   
\end{equation*}
\begin{equation*}
   \ \ \ \ \ \ \ \ \ - H\int {\frac {\partial \ln \chi} {\partial x} \bigg (h_{\oplus}  (\ddot{h}_{\otimes, 0} - h^{\prime \prime}_{\otimes,0} ) - h_{\otimes}  (\ddot{h}_{\oplus, 0} - h^{\prime \prime}_{\oplus,0})\bigg )d^4x} 
\end{equation*}
\begin{equation}
 \ \ \ \ \ \ \ \ \ + H \int{ \frac {\partial \ln \chi} {\partial t} \bigg (h_{\oplus}  (\ddot{h}_{\otimes, 1} - h^{\prime \prime}_{\otimes,1} ) - h_{\otimes}  (\ddot{h}_{\oplus, 1} - h^{\prime \prime}_{\oplus,1})\bigg )d^4x}   
\end{equation}
         and,
\begin{equation}
\mathcal{A}_\phi = \int {\bigg [ \frac {a_1} {\chi} \eta^{\mu \nu} \chi_{ ,\mu} \chi_{ ,\nu} + a_2 \chi - \frac {a_1} {\chi} \bigg ( h_\oplus (\chi_{,2}\chi_{,2} - \chi_{,3}\chi_{,3})
 + 2 h_\otimes \chi_{,2}\chi_{,3}\bigg ) \bigg] d^4x} \ \ .    
\end{equation}         
    
   After defining  $\psi (x^\mu)  \equiv \ln {(\chi (x^\mu))} $, the \lowercase{DYNAMICAL EQUATIONS OF MOTION THAT ENSUE FROM} variations of $\mathcal{A}$ (eq.(5.2)) with respect to $ h_{\oplus}$, $h_{\otimes}$ and $\psi $ are given by,
 \begin{equation*}
 \ddot{h}_\oplus - h^{\prime \prime}_\oplus= -\frac{8 \pi}{m^2_{Pl}} \bigg [ H \bigg \lbrace \frac {\partial^2 \psi}{\partial t \partial x} (
\ddot{h}_{\otimes} + h^{\prime \prime}_{\otimes}) - h_{\otimes,0 1}\bigg (\frac{\partial^2 \psi} {\partial t^2} + \frac{\partial^2 \psi} {\partial x^2}\bigg ) 
+ \frac {\partial \psi}{\partial x} (\ddot{h}_{\otimes, 0} - h^{\prime \prime}_{\otimes,0})  - \frac {\partial \psi}{\partial t} (
\ddot{h}_{\otimes, 1} - h^{\prime \prime}_{\otimes,1}) \bigg \rbrace +     
 \end{equation*}  
 \begin{equation}
 \ \ \ \ \ \ \ \ \ +  a_1 \exp{(\psi)}\bigg \lbrace \bigg (\frac {\partial \psi}{\partial y} \bigg )^2 - \bigg (\frac {\partial \psi}{\partial z} \bigg )^2 \bigg \rbrace \bigg]\ ,   
 \end{equation}  
\begin{equation*}
\ddot{h}_\otimes - h^{\prime \prime}_\otimes= \frac{8 \pi}{m^2_{Pl}} \bigg [ H \bigg \lbrace \frac {\partial^2 \psi}{\partial t \partial x} (
\ddot{h}_{\oplus} + h^{\prime \prime}_{\oplus}) - h_{\oplus,0 1}\bigg (\frac{\partial^2 \psi} {\partial t^2} + \frac{\partial^2 \psi} {\partial x^2}\bigg ) 
+ \frac {\partial \psi}{\partial x} (\ddot{h}_{\oplus, 0} - h^{\prime \prime}_{\oplus,0})  - \frac {\partial \psi}{\partial t} (
\ddot{h}_{\oplus, 1} - h^{\prime \prime}_{\oplus,1}) \bigg \rbrace +     
\end{equation*}    
\begin{equation}
 \ \ \ \ \ \ \ \ \ \  + 2 a_1 \exp{(\psi)}\frac {\partial \psi}{\partial y} \frac {\partial \psi}{\partial z}  \bigg]   
\end{equation}
 \lowercase{AND}
 \begin{equation}
\partial^\mu \partial_\mu \psi + \frac {1} {2} \eta^{\mu \nu} \frac {\partial \psi}{\partial x^\mu} \frac {\partial \psi}{\partial x^\nu} - \frac {a_2} {4 a_1} = h_\oplus \bigg [\frac {\partial^2 \psi}{\partial y^2} - \frac {\partial^2 \psi}{\partial z^2} + \frac {1} {2} \bigg (\frac {\partial \psi}{\partial y}\bigg )^2
-  \frac {1} {2} \bigg (\frac {\partial \psi}{\partial z} \bigg )^2 \bigg ] +2 h_\otimes \bigg [\frac {\partial^2 \psi}{\partial y \partial z} + \frac {1} {2} \frac {\partial \psi}{\partial y} \frac {\partial \psi}{\partial z} \bigg ] \ \ .     
 \end{equation}

(T\lowercase{HE DERIVATIVES WITH RESPECT TO TIME $t$ AND X-COORDINATE ARE DENOTED BY DOT AND PRIME, RESPECTIVELY.})
        
 A\lowercase{ SIMPLIFICATION OCCURS IF WE ASSUME $\psi $ TO DEPEND ONLY
   ON THE X-COORDINATE AND TIME, LIKE THE} GW amplitude itself.   
  W\lowercase{ITH $\psi = \psi (t,x)$, THE LAST TERMS IN EQS.(5.5) AND (5.6) DROP OUT.} I\lowercase{NTRODUCING  THE  CIRCULARLY POLARIZED} GW \lowercase{AMPLITUDES, AS DEMONSTRATED IN THE REFERENCE [21],}
    $$h_R = \frac {1} {\sqrt {2}} (h_\oplus + i h_\otimes)\ \ \ \mbox{and} \ \ \ h_L = \frac {1} {\sqrt {2}} (h_\oplus - i h_\otimes) \ ,$$
\lowercase{ONE FINDS THAT THE COUPLED DIFFERENTIAL EQUATIONS  GIVEN BY   EQS.(5.5) AND (5.6), INVOLVING $h_\oplus $ AND $h_{\otimes}$,   GET MUTUALLY SEPARATED,}
\begin{equation}
 \ddot{h}_R - h^{\prime \prime}_R= -\frac{8 \pi H i}{m^2_{Pl}} \bigg [\dot{\psi} \frac {\partial} {\partial x} (\ddot{h}_R - h^{\prime \prime}_R)
 - \psi^\prime  \frac {\partial} {\partial t} (\ddot{h}_R - h^{\prime \prime}_R) + \dot {h}^\prime_R (\ddot{\psi} + \psi^{\prime \prime}) - 
 \dot{\psi}^\prime (\ddot{h}_R + h^{\prime \prime}_R) \bigg ]   
\end{equation}
\lowercase{ AND}
\begin{equation}
 \ddot{h}_L - h^{\prime \prime}_L= \frac{8 \pi H i}{m^2_{Pl}} \bigg [\dot{\psi} \frac {\partial} {\partial x} (\ddot{h}_L - h^{\prime \prime}_L)
 - \psi^\prime  \frac {\partial} {\partial t} (\ddot{h}_L - h^{\prime \prime}_L) + \dot {h}^\prime_L (\ddot{\psi} + \psi^{\prime \prime}) - 
 \dot{\psi}^\prime (\ddot{h}_L + h^{\prime \prime}_L) \bigg ]  \ ,  
\end{equation}
displaying a symmetry under $R \longleftrightarrow L$ in the above equations modulo a negative sign, corresponding to the two independent  circular polarizations.
     
  S\lowercase{INCE $\psi = \psi (t,x)$, THE EQUATION OF MOTION FOR THE FOUR-FORM GIVEN BY EQ.(5.7) SIMPLIFIES TO,}
\begin{equation}
\ddot{\psi} - \psi^{\prime \prime} + \frac {1} {2} (\dot{\psi}^2 - \psi^{\prime\ 2}) + 2 \mu^2=0    
\end{equation}  
   where $\mu \equiv \sqrt{- \frac {a_2} {4 a_1} }$ is akin to  the rest mass of the four-form field.
   
 S\lowercase{EEKING EXACT SOLUTIONS OF EQS.(5.8) AND (5.10), INVOLVING MONOCHROMATIC AND CIRCULARLY POLARIZED} GWs as well as $\psi $, we substitute,
\begin{equation*}
 h_R(t,x)=h \exp(i(\omega t- kx))   
\end{equation*} 
\lowercase{IN EQ.(5.8), LEADING TO A RELATION,}
\begin{equation}
\omega^2 - k^2= -\beta [k (\omega^2 - k^2) \dot{\psi} + \omega (\omega^2 - k^2) \psi^\prime - i \omega k (\ddot{\psi} + \psi^{\prime \prime})
 - i (\omega^2 + k^2) \dot{\psi}^\prime ]    
\end{equation}
 where $\beta \equiv \frac{8 \pi H }{m^2_{Pl}}$.

I\lowercase{F $w=\pm k$, EQ.(5.11) IMPLIES,}
\begin{equation}
2\dot{\psi}^\prime \pm  (\ddot{\psi} + \psi^{\prime \prime})=0 \ \ .    
\end{equation}
The only self-consistent solution of eqs.(5.10) and (5.12) is,
\begin{equation}
 \psi(t,x)= (b_0 - \frac {\mu^2} {b_0}) t + (b_0 + \frac {\mu^2} {b_0}) x    
\end{equation}
\lowercase{WHERE $b_0$ IS AN INTEGRATION CONSTANT.} B\lowercase{UT THE ABOVE SOLUTION IS UNPHYSICAL, AS IT  IMPLIES
 AN EXPONENTIALLY GROWING $\chi (t,x) $.}
 On the other hand, when $w^2 > k^2$,
 \begin{equation}
k \dot{\psi} + \omega \psi^\prime + \frac {1} {\beta} =0 \Rightarrow \psi^{\prime \prime} = \frac {k^2} {w^2}\  \ddot{\psi}     
 \end{equation}
\lowercase{AND THEREFORE,} $\psi = \psi (x - \frac {\omega} {k} t)$, leading to an exact solution for $b_1 \geq \lambda$,
 \begin{equation}
\psi= \ln {\bigg (\sqrt{\frac {b_1} {\lambda}}\cos^2 \bigg (\pm \sqrt{\frac{\lambda} {2}} (t - \frac {k} {\omega}  x) + b_2 \bigg ) \bigg )}     
 \end{equation}
where $\lambda \equiv \frac{2 \omega^2 \mu^2} {\omega^2 - k^2}$, while \lowercase{$b_1$ and $b_2$ ARE INTEGRATION CONSTANTS.}

 Hence,
 \begin{equation}
 \chi  (t - \frac {k} {\omega}  x) =\exp (\psi) \propto \cos^2 \bigg (\pm \sqrt{\frac{\lambda} {2}} (t - \frac {k} {\omega}  x) + b_2 \bigg )\ ,    
 \end{equation}
 with $| \omega | > | k |$, is physically meaningful and is an 
acceptable solution.

I\lowercase{NDEED, WE FIND THAT IT IS POSSIBLE TO HAVE EXACT SOLUTIONS CORRESPONDING TO THE FOUR-FORM AND CIRCULARLY POLARIZED, MONOCHROMATIC} GWs \lowercase{WITH PHASE VELOCITY EXCEEDING THE SPEED OF LIGHT.}
The caveat, however, is that \lowercase{EQ.(5.10) BEING A NONLINEAR DIFFERENTIAL EQUATION, DERIVING EXACT SOLUTIONS CORRESPONDING TO} eqs.(5.8) and (5.10)  \lowercase{WITH SUPERPOSED} GW \lowercase{MODES WITH A PHYSICALLY ACCEPTABLE GROUP VELOCITY IS NON-TRIVIAL.}

\vskip 1.5 em
\section{Conclusions}
\vskip 1.5 em

Based on the invariances of the Minkowski tensor and the flat space-time Levi-Civita tensor under proper Lorentz transformations, the present study explores the possibility of extending Einstein's geometrical formulation of gravitation by including another geometrical  field, $\tilde {w} $ -   a generalisation of the Levi-Civita symbol, in the theory. The geometrodynamical four-form $\tilde{w} $ leads not only to a dynamical torsion, it also generates Chern-Simon (CS) extensions in the 3+1-dimensional space-times.  Torsion, among several other interesting implications, is also important in teleparallel gravity theories \cite{Hehl1, TelepGravity}.  

Pursuing closely the seminal work of Jackiw and Pi \cite{Jackiw} but avoiding the use of an unphysical Lorentz  vector,  CS coupling between electromagnetic fields and $\tilde{w}$ comes about naturally, bringing about  a modification of Einstein-Maxwell equations. Adopting  the formalism of \cite{Jackiw},  a  gravitational CS term is also constructed that leads to a modified Cotton tensor. However, when the dynamical equation of the four-form is used  in  the ensuing  Einstein equation,  it is shown that it is  the standard Cotton tensor that  affects the space-time dynamics.

The scalar-density $\phi $  associated with  $\tilde {w} $ leads to a well-defined exterior derivative that turns a differential p-form density into  a (p+1)-form density of same  weight.  Since the notion of an n-form and exterior derivative in a differential manifold does not require either an affine connection or a metric,  further studies are required to investigate the role of $\tilde {w} $ in situations where metric is degenerate as well as its impact on the  manifold-orientability.

The geometrodynamical four-form field is shown to correspond to a pseudo-scalar particle. From a simplified semi-classical treatment, it is demonstrated that the Bose-Einstein condensates of such pseudo-scalars can give rise to formation of supermassive black holes through an interplay of self-gravity and quantum mechanics. Therefore, self-gravitating BECs of pseudo-scalar particles associated with the geometrodynamical four-form field may solve the long standing  problem concerning the frequent discoveries of tens of  billion solar mass  SMBHs at epochs when the universe was less than  billion years old.

It  is also  proved that \lowercase{
   IN THE PRESENCE OF THE DYNAMICAL FOUR-FORM, EQUATIONS FOR THE LINEARLY POLARIZED GRAVITATIONAL WAVES  CAN BE DECOUPLED USING CIRCULARLY POLARIZED WAVEFORMS AS WELL AS   EXACT AND  SELF-CONSISTENT SOLUTIONS OF THE DYNAMICAL FOUR-FORM EQUATION COUPLED TO 
  A MONOCHROMATIC GRAVITATIONAL WAVE (ALBEIT WITH PHASE VELOCITY $>c$) 
  CAN BE OBTAINED}. However, much more work is needed  to obtain realistic  solutions of gravitational waves, with group  velocities $\leq c$, propagating in the background consisting of $\tilde{w}$.  

\section{Acknowledgements}

It is a pleasure  to thank  Prof. Vesselin Petkov for organising the Third Minkowski Meeting that entailed stimulating discussions on the fundamental aspects of space-time physics.

\end{document}